\documentclass[pra,showpacs,showkeys,twocolumn]{revtex4-1}

\usepackage{amsmath}
\usepackage{amsfonts}            
\usepackage{amssymb}    

\usepackage[mathcal,mathscr]{eucal}

\usepackage{mathrsfs}

\usepackage{eufrak}

\usepackage{graphicx}
\usepackage{epstopdf}

\usepackage{dcolumn}
\usepackage{bm}

\makeatletter
\def\@dotsep{4.5}
\makeatother

\begin{document}


\title{Shaping interactions between polar molecules with  far-off-resonant light}

\author{Mikhail Lemeshko}

\email{mikhail.lemeshko@gmail.com}

\affiliation{%
Fritz-Haber-Institut der Max-Planck-Gesellschaft, Faradayweg 4-6, D-14195 Berlin, Germany
}%




\date{\today}

\begin{abstract}

We show that dressing polar molecules with a far-off-resonant optical field leads to new types of intermolecular potentials, which undergo a crossover from the inverse-power to oscillating behavior depending on the intermolecular distance, and whose parameters can be tuned by varying the laser intensity and wavelength. We present analytic expressions for the potential energy surfaces, thereby providing direct access to the parameters of an optical field required to design intermolecular interactions experimentally.

\end{abstract}


\maketitle

An intense far-off-resonant laser field induces a retarded dipole-dipole interaction between atoms or molecules, which is of long range character and falls off as $1/r$, $1/r^2$, or $1/r^3$, depending on the separation $r$ and  the light wavevector~\cite{CraigThiruBook, ThiruMolPhys80}. This interaction is highly controllable and  brings about peculiar effects in atomic  Bose condensates, such as ``gravitational self-binding,'' rotons, and density modulations leading to a supersolid-like behavior~\cite{ODellPRL00, GiovanazziPRL02, ODellPRL03}. In contrast to atoms, molecules possess anisotropic polarizability and rotational structure, which renders laser-induced interactions more complex, and thereby offers new riches in the few- and many-body physics of field-dressed ultracold gases.

In this contribution we study the interaction between two polar and polarizable molecules in a far-off-resonant laser field. We demonstrate that an optical field gives rise to new types of intermolecular potentials, which exhibit a crossover from the inverse-power decay at short distances to an oscillating long-range behavior, and whose parameters can be  altered by tuning the laser intensity and frequency. Furthermore, for a wide range of field intensities the problem can be described by an exactly solvable two-level model, which leads to simple analytic expressions for the effective potential energy surfaces. The theory is exemplified by  optically-induced interactions between $^{40}$K$^{87}$Rb molecules, widely employed in experiments with ultracold polar gases~\cite{NiJinYeNature2010}.

We consider two identical diatomic molecules, $1$ and $2$, with a dipole moment $d$ and polarizability components, $\alpha_\parallel$ and $\alpha_\perp$, parallel and perpendicular to the molecular axis. 
In a far-off-resonant radiative field of intensity $I$, the rotational levels of each molecule undergo a dynamic Stark shift, as given by the Hamiltonian~\cite{CraigThiruBook}
\begin{equation}
	\label{Hi}
	H_{1,2} = B \mathbf{J}^2 - \frac{I}{2 c \varepsilon_0} e_j e_l^\ast \alpha^{\text{lab}, (1,2)}_{jl}(k),
\end{equation}
with $B$ the rotational constant, and $\alpha^\text{lab}_{jl}(k)$ the dynamic polarizability tensor in the laboratory frame. We assume a laser beam propagating along the positive $Y$ direction with the wavevector $\mathbf{k} = k \mathbf{\hat{Y}}$, and linear polarization along the $Z$ axis, $\mathbf{\hat{e}}=\mathbf{\hat{Z}}$. Given that the only nonzero polarizability components in the molecular frame are $\alpha_{zz}=\alpha_\parallel$ and $\alpha_{xx}=\alpha_{yy}=\alpha_\perp$, and using $B$ as a unit of energy, Hamiltonian~(\ref{Hi}) can be recast as:
\begin{equation}
	\label{Halpha}
	H_{1,2} =  \mathbf{J}^2   - \Delta \eta (k) \cos^2 \theta_{(1,2)} - \eta_\perp (k),
\end{equation}
where $\theta_{(1,2)}$ is the angle between the molecular axis and the polarization vector of the laser field. The dimensionless interaction parameter $\Delta \eta(k) =  \eta_\parallel(k) - \eta_\perp (k)$ with  $\eta_{\parallel,\perp} (k) =  \alpha_{\parallel, \perp} (k) I /(2\varepsilon_0 c B)$. We note that eq.~(\ref{Halpha}) was derived in Refs.~\cite{FriHerPRL95, FriHerJPC95} using the semiclassical approach and the rotating wave approximation.  All rotational levels exhibit a constant shift of $\eta_\perp$,  given by the second term of eq.~(\ref{Halpha}), which will be omitted hereafter.

The polarization vector of an optical field defines an axis of cylindrical symmetry, $Z$. The projection, $M$, of the angular momentum $\mathbf{J}$ on $Z$ is then a good quantum number, while $J$ is not. However, one can use the value of $J$ of the field-free rotational state, $Y_{J, M} (\theta, \phi)$, that adiabatically correlates with the hybrid state as a label, designated by $\tilde{J}$, so that $|\tilde{J}, M; \Delta \eta \rangle \to Y_{J, M}$ for $\Delta \eta \to 0$. For emphasis, we also label the values of $\tilde{J}$ by tilde, e.g.\ with $\tilde{0}$ corresponding to $\tilde{J}=0$.

Induced-dipole interaction~(\ref{Halpha}) preserves parity, hybridizing states with even or odd $J$'s,
\begin{equation}
	\label{PendularState}
	|\tilde{J}, M; \Delta \eta \rangle = \sum_{J} c_{J M}^{\tilde{J}, M} (\Delta \eta ) Y_{J M} , \hspace{0.2cm} J+\tilde{J} \hspace{0.15cm} \text{even},
\end{equation}
and therefore aligns molecules in the laboratory frame. Aligned molecules possess no space-fixed dipole moment, in contrast to species oriented by an electrostatic field.

We note that at the far-off-resonant wavelengths usually employed in alignment and trapping experiments ($\sim$1000 nm), the dynamic polarizability $\alpha_{ij} (k)$ approaches its static limit, $\alpha_{ij} (0)$, for a number of molecules, e.g.\ CO, N$_2$, and OCS. However, this is not the case for alkali dimers having low-lying excited $^1\Sigma$ and $^1\Pi$ states, such as KRb and RbCs. Virtual transitions to these states contribute to the ground-state dynamic polarizability, rendering it a few times larger than the static value~\cite{KotochigovaDeMille10, DeiglmayrDulieuJCP08}.

\begin{figure}[t]
\includegraphics[width=7.7cm]{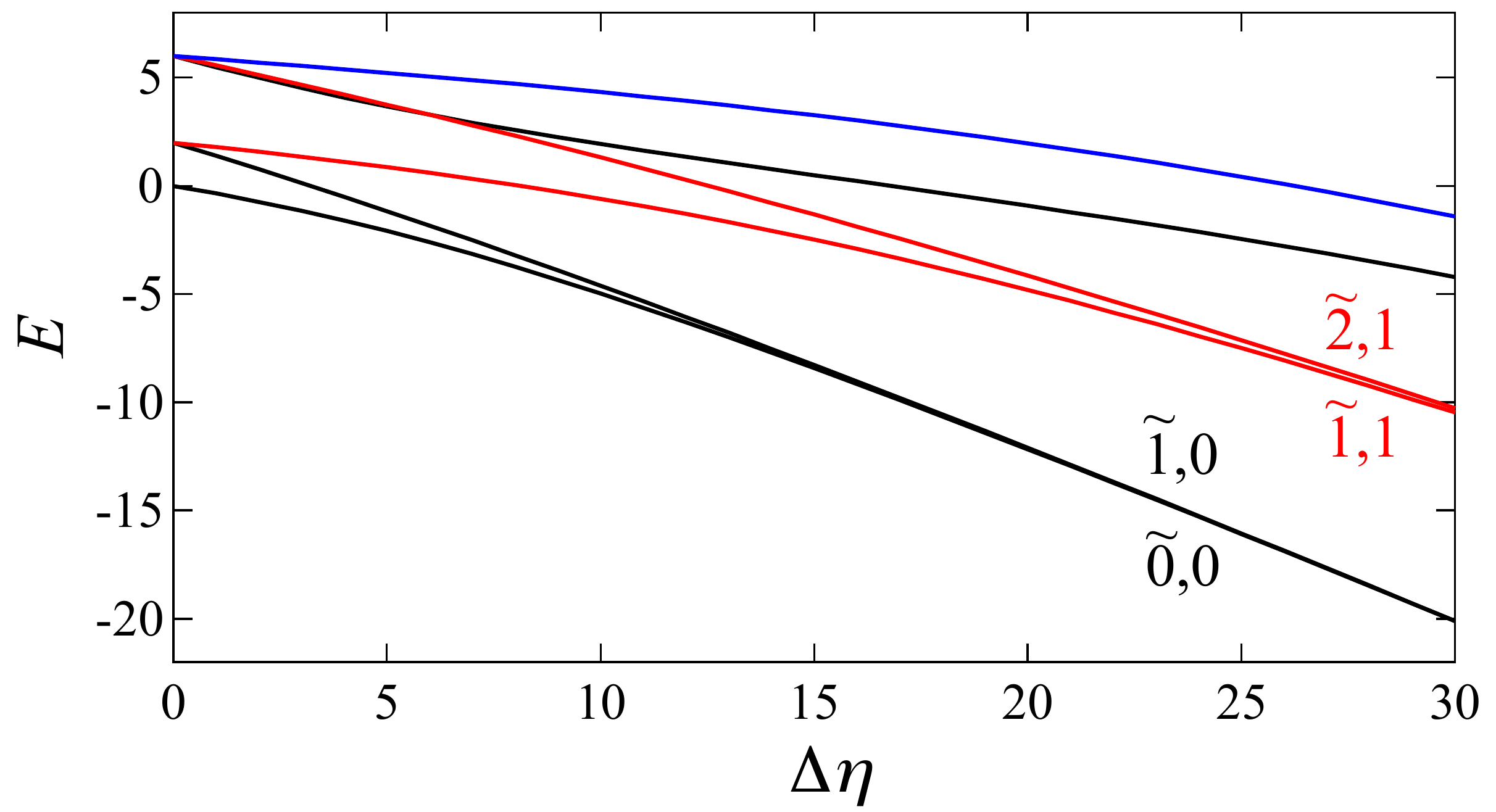}
\caption{\label{fig:AC_Stark} Lowest rotational energy levels of a molecule in a far-off-resonant laser field, depending on the field-strength parameter $\Delta \eta$. Different colors correspond to $M=0$ (black), 1 (red), and 2 (blue). Energy is in units of $B$, two lowest tunneling doublets are labeled as $\tilde{J}, M$. For $\Delta \eta \gtrsim 15$ the splitting of the lowest tunneling doublet can be accurately estimated as $\Delta E = |E| \exp (3.6636 - 2\sqrt{\Delta \eta})$, with $E=2 \sqrt{\Delta \eta} - \Delta \eta - 1$~\cite{FriHerJPCA99}.}
\end{figure}

Figure~\ref{fig:AC_Stark} illustrates the dynamic Stark effect on rotational levels of a diatomic molecule. A far-off-resonant optical field of sufficiently large intensity leads to formation of ``tunneling doublets" -- closely lying states of opposite parity with the same $M$ and $\Delta \tilde{J} = 1$~\cite{FriHerPRL95}. The doublet states can be mixed by extremely weak electrostatic fields, leading to strong molecular orientation in the laboratory frame~\cite{FriHerJCP99, FriHerJPCA99}. The energy gap between neighboring doublets increases with the field intensity, and is proportional to $2\sqrt{\Delta\eta}$ in the strong-field limit~\cite{HaerteltFriedrichJCP08}. As we demonstrate below, at large $\Delta\eta$ interaction between two ground-state molecules can be described within the lowest tunneling doublet.

In the absence of fields, two polar molecules interact via the dipole-dipole interaction:
\begin{equation}
	\label{DDpot}
	V_\text{dd} (r) = \frac{\hat{\mathbf{d}}^{(1)}_j \hat{\mathbf{d}}^{(2)}_l}{r^3} (\delta_{jl} - 3 \hat{\mathbf{r}}_j \hat{\mathbf{r}}_l),
\end{equation}
where $\mathbf{\hat{d}}^{(1,2)}=\mathbf{d}^{(1,2)}/d$ are unit dipole moment vectors of the molecules, $\hat{\mathbf{r}}$ is the unit vector defining the intermolecular axis, and 
\begin{equation}
	\label{r0}
	r_0 = \left( \frac{d^2}{4 \pi \varepsilon_0 B} \right)^{1/3}
\end{equation}
is introduced as a unit of length. In the field-free case, interaction between two ground-state polar molecules has an isotropic asymptotic behavior, $V_\text{dd}(r) = - 1/(6 r^6)$.

Far-off-resonant laser light interacts with molecular polarizability, thereby inducing oscillating dipole moments on each of the two molecules. Retarded interaction between these instantaneous dipoles leads to an additional term in the intermolecular potential~\cite{CraigThiruBook, ThiruMolPhys80, SalamPRA07},
\begin{equation}
	\label{VlaserInd}
	V_{\alpha \alpha} (k, \mathbf{r}) = \frac{I}{4 \pi \varepsilon_0^2 c} e_i^\ast \alpha^{\text{lab},1}_{ij}(k) V_{jl}(k,\mathbf{r}) \alpha^{\text{lab},2}_{l n}(k) e_n \cos(\mathbf{k r}),
\end{equation}
with $ V_{jl}$ the retarded dipole-dipole interaction tensor,
\begin{multline}
	\label{Vjk}
	V_{jl}(k,\mathbf{r}) = \frac{1}{ r^3} \bigl [ (\delta_{jl} - 3 \hat{\mathbf{r}}_j  \hat{\mathbf{r}}_l ) (\cos kr + kr \sin kr) \\ -  (\delta_{jl} - \hat{\mathbf{r}}_j  \hat{\mathbf{r}}_l ) k^2 r^2 \cos kr \bigr ]
\end{multline}

For the laser light linearly polarized along the $Z$-axis, optically-induced dipole-dipole interaction~(\ref{VlaserInd}) can be rewritten in dimensionless form,
\begin{equation}
	\label{VlaserIndred}
	V_{\alpha \alpha} (k, \mathbf{r}) = \frac{\Delta \eta (k)}{\xi(k)}  \tilde{\alpha}^{\text{lab},1}_{Z j}(k) V_{jl}(k,\mathbf{r}) \tilde{\alpha}^{\text{lab},2}_{l Z}(k) \cos(\mathbf{k r}),
\end{equation}
where  energy is measured in units of $B$, distance in units of $r_0$, and $k$ in units of $r_0^{-1}$, and $\tilde{\alpha}_{ij} = \alpha_{ij}/\Delta \alpha$ with $\Delta \alpha = \alpha_\parallel - \alpha_\perp$ is the reduced polarizability tensor. The dimensionless parameter,
\begin{equation}
	\label{Xiparam}
	\xi(k) = \frac{d^2}{2 \Delta \alpha(k) B},
\end{equation}
characterizes the relative strength of the permanent-dipole and induced-dipole interactions and is on the order of $10^2-10^3$ for polar alkali dimers. We note that in eq.~(\ref{VlaserInd}) we neglected a ``static'' term due to the coupling of the dipole moment of one molecule with the hyperpolarizability of another via the optical field~\cite{BradshawPRA05, SalamPRA07}. This interaction is independent of $k$ and only becomes comparable to the dipole-dipole potential (\ref{DDpot}) at much larger intensities ($I \gtrsim 10^{14}$ W/cm$^2$).


In the Born-Oppenheimer  approximation, effective interaction potentials $V_\text{eff}(\mathbf{r})$ are obtained by diagonalizing the Hamiltonian,
\begin{equation}
	\label{Hamil}
	H = H_1 + H_2 + V_\text{dd} + V_{\alpha \alpha}, 
\end{equation}
for fixed intermolecular separations $\mathbf{r} = (r, \theta, \phi)$. In dimensionless units, the dipole-dipole potential~(\ref{DDpot}) is on the order of unity, while the strength of the optically-induced interaction~(\ref{VlaserIndred}) is given by $\Delta \eta/\xi$. Therefore, for the field strength parameter satisfying the inequality $1\ll \sqrt{\Delta \eta} \ll \xi(k)$, both interaction terms are much smaller than the energy gap between neighboring tunneling doublets. Hence, in the basis of field-dressed states, $\vert \tilde{J}_1 M_1, \tilde{J}_2 M_2 \rangle$, the interaction between two ground-state molecules can be treated within the lowest tunneling doublet, $\vert \tilde{0}0, \tilde{0} 0 \rangle$--$\vert \tilde{1}0, \tilde{1} 0 \rangle$. Given that the optically-induced potential~(\ref{VlaserIndred})  and the dipole-dipole interaction (\ref{DDpot}) mix only states of the same and the opposite parity, respectively~\cite{CrossGordonJCP66, LemeshkoOpticalLong11}, the Hamiltonian matrix takes the form:
\begin{equation}
	\label{Hmatr}
	H_\text{rel} =  \left( \begin{array}{cc} U_{\alpha \alpha}^{\tilde{0}} & U_\text{dd} \\  U_\text{dd} & U_{\alpha \alpha}^{\tilde{1}} + 2 \Delta E  \end{array} \right),
\end{equation}
where $\Delta E (\Delta \eta)$ is the splitting between the tunneling-doublet levels $\vert \tilde{0}, 0 \rangle$ and $\vert \tilde{1}, 0 \rangle$, cf. Fig.~\ref{fig:AC_Stark},  and the matrix elements are given by:
\begin{equation}
	\label{Udd}
	U_\text{dd} (\mathbf{r}) = \frac{1- 3 \cos^2 \theta}{r^3} G(\Delta \eta),
\end{equation}
\begin{multline}
	\label{Uind}
	U_{\alpha \alpha}^{\tilde{J}}  (\mathbf{k}, \mathbf{r}) = \frac{\Delta \eta (k)}{\xi(k)}   \frac{\cos{(\mathbf {k r})}}{r^3} K_{\tilde{J}} (\Delta \eta, \alpha/\Delta \alpha) \\
	\times \left \{ -\sqrt{\frac{2}{3}} a (kr) 	+ \left [a(kr) - 3b(kr) \right ] \sqrt{\frac{8 \pi}{15}} Y_{20} (\theta, \phi) \right  \},
\end{multline}
with $\alpha = (\alpha_\parallel + 2 \alpha_\perp)/3$ the average molecular polarizability, $a(x) = x^2 \cos x$, and $b(x) = (\cos x + x \sin x)$.

\begin{figure}[t]
\includegraphics[width=9.2cm]{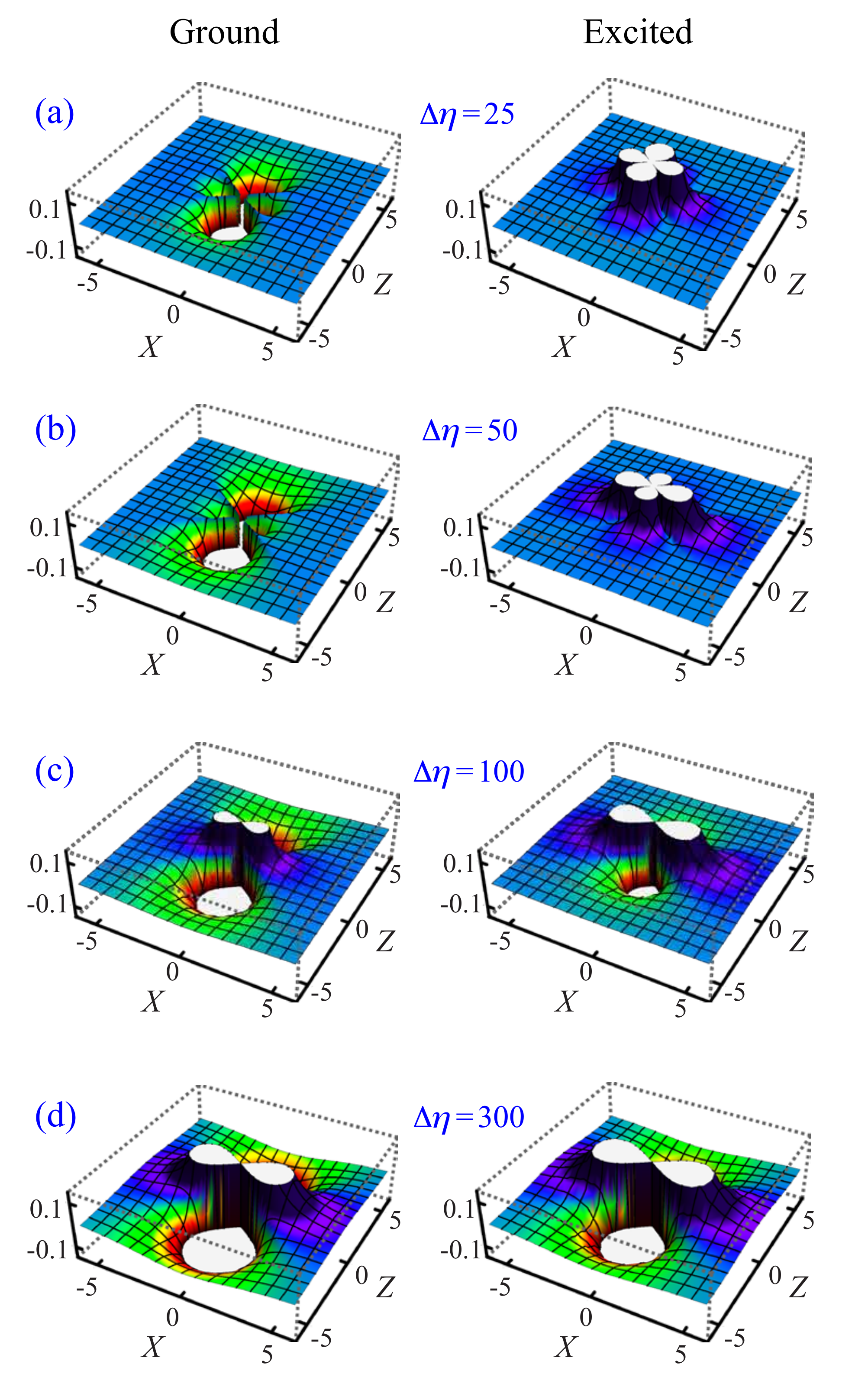}
\caption{\label{fig:KRb_pots} Short-range behavior of optically-induced KRb--KRb potentials in the XZ plane ($\phi=0$), for different values of $\Delta \eta$. Left and right columns correspond, respectively, to the ground $\vert \tilde{0} 0, \tilde{0} 0 \rangle$ state, and excited $\vert \tilde{1} 0, \tilde{1} 0 \rangle$ state  potentials, as given by eq.~(\ref{Veff}). The dependence of the short-range potentials on $\phi$ is negligible. Potential energy is in units of $B$, with $V_\text{eff}(r \to \infty)$ chosen as zero. The laser beam propagates along the $Y$ axis, $\mathbf{k} \parallel \hat{\mathbf{Y}}$, with the polarization $\hat{\mathbf{e}} \parallel \hat{\mathbf{Z}}$. See text.}
\end{figure}

The factors $G(\Delta \eta)$ and $K_{\tilde{J}} (\Delta \eta, \alpha/\Delta \alpha)$ have an analytic representation in terms of the coefficients $c_{J 0}^{\tilde{J}, 0} (\Delta \eta )$ of eq.~(\ref{PendularState})~\cite{LemeshkoOpticalLong11}. They are on the order of unity and can be analytically estimated in the strong-field limit, $\Delta \eta \to \infty$, with good accuracy:
\begin{equation}
	\label{Gstrong}
	G(\Delta \eta) = \left[ 1 - \Delta \eta^{-1/4} F \left (\tfrac{1}{2}\Delta \eta^{-1/4} \right) \right]^2,
\end{equation}
\begin{equation}
	\label{Kstrong}
	K_{\tilde{J}=0,1} (\Delta \eta) =  \sqrt{\frac{2}{3}}  \left[ \frac{\alpha_\parallel}{\Delta \alpha} - \Delta \eta^{-1/4} F \left (\Delta \eta^{-1/4} \right) \right]^2,
\end{equation}
where $F(x) = \exp (-x^2) \int_0^x \exp (y^2) dy$ is the Dawson integral~\cite{AbramowitzStegun}.

In such a way, the effective potentials are given by solutions of (\ref{Hmatr}):
\begin{multline}
	\label{Veff}
	V_\text{eff} (\mathbf{k}, \mathbf{r}) = \Delta E + \frac{U_{\alpha \alpha}^{\tilde{0}} (\mathbf{k}, \mathbf{r}) + U_{\alpha \alpha}^{\tilde{1}} (\mathbf{k}, \mathbf{r})}{2} \\
	\pm \frac{\sqrt{4 U_\text{dd}^2 (\mathbf{r})  + \left[ U_{\alpha \alpha}^{\tilde{1}} (\mathbf{k}, \mathbf{r}) - U_{\alpha \alpha}^{\tilde{0}} (\mathbf{k}, \mathbf{r}) + 2 \Delta E \right]^2}}{2},
\end{multline}
with the minus sign corresponding to the ground $\vert \tilde{0} 0, \tilde{0} 0 \rangle$ state effective potential.

Figure~\ref{fig:KRb_pots} shows the short-range behavior of effective potentials~(\ref{Veff}), induced between two $^{40}$K$^{87}$Rb  molecules by a laser field of wavelength $\lambda=1090$~nm ($k=0.021/r_0$), corresponding to the following values of parameters: $\xi = 96.5$, $\alpha / \Delta \alpha = 0.62$, and $r_0 = 36.1$~\AA~\cite{AldegundePRA08, KotochigovaDeMille10, DeiglmayrDulieuJCP08, AymarDulieuJCP05}.  

The behavior of the effective potentials is dictated by the interplay between static dipole-dipole and optically-induced dipole-dipole interactions, eqs.~(\ref{DDpot}) and (\ref{VlaserInd}). At small distances, $kr \ll 1$, the dominant contribution to eq.~(\ref{Veff}) reads:
\begin{equation}
	\label{VeffShort}
	V_\text{eff} (kr \ll 1) \approx  \frac{\vert 1- 3\cos^2 \theta \vert}{r^3} \left[ \sqrt{\frac{3}{2}} \frac{\Delta \eta}{\xi} s(\theta) \pm G(\Delta \eta) \right],
\end{equation}
where $s(\theta) = \text{sgn} [1-3\cos^2\theta]$. Eq.~(\ref{VeffShort}) reveals the presence of a critical value, $\Delta \eta_c = \xi \sqrt{2/3} G(\Delta \eta)$, that determines the sign of the short-range potential  ($\Delta \eta_c \sim 65$ for KRb). For $\Delta \eta < \Delta \eta_c $, eq.~(\ref{VeffShort}) is governed by the second term in the square brackets, and  the potentials are always attractive in the ground state and repulsive in the excited state for any angle $\theta$, except for $\theta = \pm \arccos (\pm 3^{-1/2})$, where $V_\text{eff} (kr \ll 1)$ vanishes identically, cf. Fig~\ref{fig:KRb_pots} (a), (b). This behavior is qualitatively different from the dipole-dipole interaction between two polar molecules oriented along the $Z$ axis,  $V_{dd}^{\uparrow \uparrow}=(1 - 3\cos^2 \theta)/r^3$, whose sign alternates in dependence on $\theta$. On the other hand, for  $\Delta \eta  > \Delta \eta_c $, the sign of $V_\text{eff}$ becomes angle-dependent, due to the interplay between the terms in the square brackets of eq.~(\ref{VeffShort}), resulting in the behavior similar to $V_{dd}^{\uparrow \uparrow}$: the potential is attractive at $\theta=0, \pi$ and repulsive at $\theta=\pi/2$, cf. Fig.~\ref{fig:KRb_pots} (c), (d). Both below and above $\Delta \eta_c$ the inverse-power decay rate of $V_\text{eff}$ can be tuned by changing $\Delta \eta$.


\begin{figure}[t]
\includegraphics[width=9.4cm]{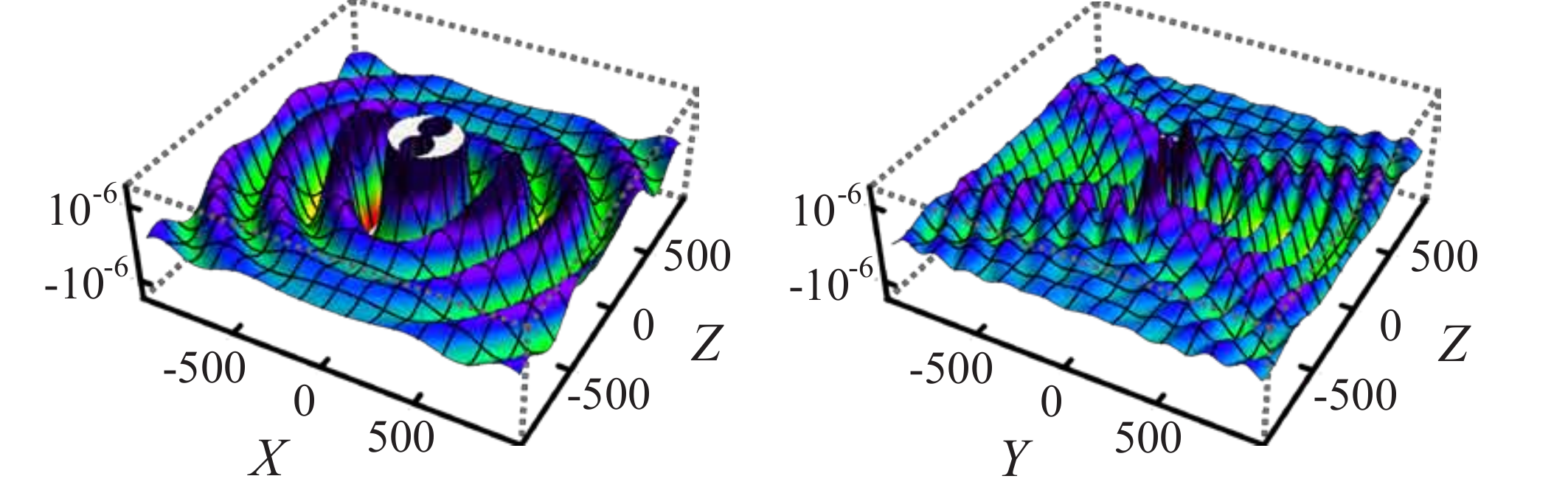}
\caption{\label{fig:KRb_potsLong} Long-range behavior of the ground $\vert \tilde{0} 0, \tilde{0} 0 \rangle$ state potentials in the $XZ~(\phi=0)$ and $YZ~(\phi=\pi/2)$ planes, for $\Delta \eta =100$. Long-range behavior is similar for the excited $\vert \tilde{1} 0, \tilde{1} 0 \rangle$ state, the magnitude of the oscillations scales with $\Delta \eta$, as given by eq.~(\ref{Vefflong}). Potential energy is in units of $B$, with $V_\text{eff}(r \to \infty)$ chosen as zero. The laser beam propagates along the $Y$ axis, $\mathbf{k} \parallel \hat{\mathbf{Y}}$, with the polarization $\hat{\mathbf{e}} \parallel \hat{\mathbf{Z}}$.  See text.}
\end{figure}

At large distances, $kr \gg 1$, the optically-induced potential~(\ref{Uind}) is proportional to $1/r$ and hence dominates over the dipole-dipole interaction~(\ref{Udd}), resulting in an asymptotic behavior given by:

\begin{equation}
	\label{Vefflong}
	V_\text{eff} (kr \gg 1) \approx - k^2 \frac{\Delta \eta}{\xi}  \sqrt{\frac{3}{2}} \frac{\cos ({\mathbf{k r}}) \cos (kr)}{r} \sin^2 \theta,
\end{equation}
which manifests itself in decaying oscillations, as shown in Fig.~\ref{fig:KRb_potsLong} for the case of $\Delta \eta = 100$. We note that  the $1/r$-like interaction~(\ref{Vefflong}) is of longer range than both the van der Waals and $V_{dd}^{\uparrow \uparrow}$ potentials, which may lead to intriguing scattering properties of optically-dressed molecules.

In general, the potentials of eq.~(\ref{Veff}) depend on the azimuthal angle $\phi$, as given by the $\cos (\mathbf {k r})$ term of eqs.~(\ref{Uind}) and (\ref{Vefflong}). Although in the case of $kr \ll1$ the $\phi$-dependence is negligible, the long-range behavior is strongly anisotropic in $\phi$, cf. Fig.~\ref{fig:KRb_potsLong}. As follows from eq.~(\ref{Vefflong}), the magnitude and phase of the long-range oscillations scale with $\Delta \eta$ and $k$ respectively, and are similar for the ground and excited state.

As one can see from Fig. \ref{fig:KRb_potsLong}, optically-induced potentials exhibit concentric minima. If these potential wells are deep enough, they will support long-range bound states, whose properties are completely determined by the optical field, molecular dipole moments and polarizabilities, and are independent of the details of the intermolecular potential, in a similar way to the electrostatically-induced bound states predicted by Avdeenkov and Bohn~\cite{AvdeenkovBohnPRL03}. Looking into the properties of these states represents a challenging theoretical and computational problem.

In summary, we undertook a study of intermolecular interactions in the presence of an intense far-off-resonant optical field, and provided simple analytic expressions for the resulting potential energy surfaces. Optically-induced potentials are highly controllable and are significantly different from the dipole-dipole interaction taking place between oriented polar molecules~\cite{BaranovPhysRep08}. This could open a way to novel quantum phases of laser-dressed ultracold polar gases and new methods to control molecular collisions in the ultracold regime.

The author is indebted to Bretislav Friedrich for continuous encouragement and valuable suggestions on the manuscript; to Eugene Demler, Mikhail Lukin, and Boris Sartakov for  insightful discussions; to Svetlana Kotochigova for providing dynamic polarizabilities; and to Gerard Meijer for support.

\bibliography{References_library}
\end{document}